\documentstyle[sprocl,epsfig]{article}

\def\Journal#1#2#3#4{{#1} {\bf #2}, #3 (#4)}

\def\PLB{{\em Phys. Lett.}  B}

\def\be{\begin{equation}}
\def\ee{\end{equation}}
\def\bea{\begin{eqnarray}}
\def\eea{\end{eqnarray}}

\begin{document}

\title{ON THE RANGE OF VALIDITY OF THE QCD-IMPROVED PARTON MODEL}

\author{A. SZCZUREK}

\address{H.~Niewodniczanski Institute of Nuclear Physics,
ul.~Radzikowskiego 152,\\ Krak\'ow, Poland \\E-mail:
szczurek@alf.ifj.edu.pl}


\maketitle\abstracts{
Based on the world DIS data we extract $F_2^p - F_2^n$ as a function
of Bjorken $x$ and photon virtuality $Q^2$. The standard PDFs fail
to describe the experimental data below $Q^2 <$ 7 GeV$^2$.
The trend of the data is approximately explained by our recent
model.
}
Two methods were used to extract experimental values of
$D_G = F_2^p - F_2^n$:
\begin{eqnarray}
D_G &=& F_2^p(x,Q^2) \cdot
\left(
1 - \frac{F_2^n(x,Q^2)}{F_2^p(x,Q^2)}
\right) \; , \nonumber \\
D_G &=& 2 F_2^d(x,Q^2) \cdot
\left(
\frac{1 - \frac{F_2^n(x,Q^2)}{F_2^p(x,Q^2)}}
     {1 + \frac{F_2^n(x,Q^2)}{F_2^p(x,Q^2)}}
\right)
\; .
\end{eqnarray}
The two methods above are of course equivalent when nuclear effects
are neglected. They are equivalent also in a less obvious case when
nuclear effects are taken into account and $F_2^n/F_2^p$ is
replaced by $ 2 F_2^d / F_2^p - 1$ \cite{SU_F2p_F2n}.
As an example in Fig.1 I show the $Q^2$ dependence of $F_2^p-F_2^n$
for some selected values of Bjorken $x$. The results obtained
with the two methods agree (compare left and right column).
The theoretical results obtained with different recent PDF's
\cite{PDFs} are shown for comparison. Above $Q^2 >$ 7 GeV$^2$ the parton
model describes the experimental data well. However, the predictions
of the parton model clearly deviate from from the experimental data
below $Q^2 <$ 7 GeV$^2$, especially for Bjorken $x$ between 0.2 and
0.45. The predictions of our recent model \cite{SU_inclusive}
are shown by the thick solid line and reproduce the main trend of the data.
Because the VDM contribution to $F_2^p - F_2^n$ almost cancels
\cite{SU_F2p_F2n}, the experimental data seem to present
a real modification of the parton model at small resolution scales.
This modification can be view formally as higher twist effects
\cite{SU_F2p_F2n}.
\newpage
\begin{figure}[htb]
\begin{center}
\mbox{
\epsfysize 10.0cm
\epsfbox{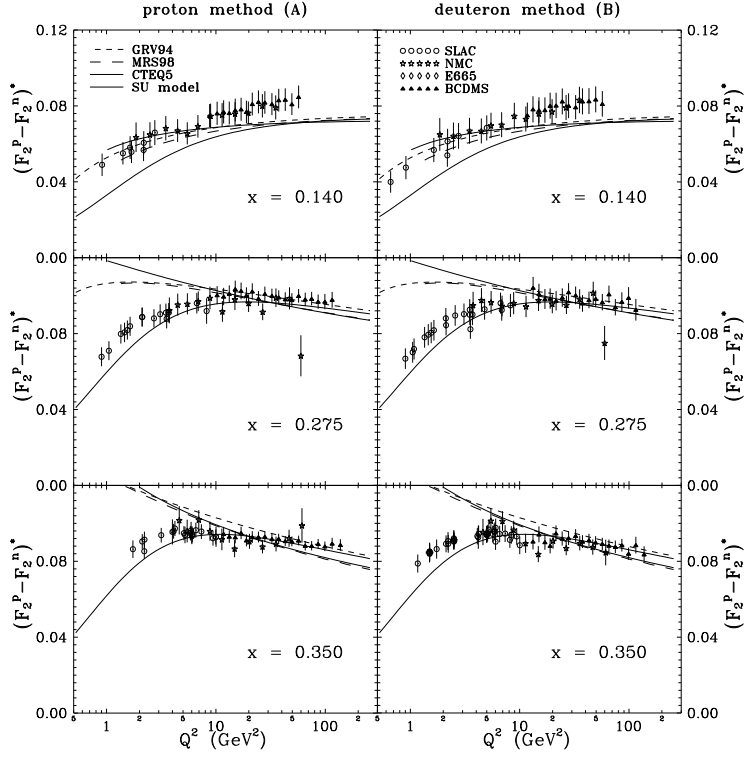}
}
\caption{\it $D_G(Q^2)$ for some selected values of Bjorken $x$.}
\end{center}
\end{figure}

\end{document}